\documentclass[%
reprint,
superscriptaddress,
 amsmath,amssymb,
 aps,
prb,
]{revtex4-2}

\usepackage{graphicx}
\usepackage{dcolumn}
\usepackage{bm}
\usepackage{here}
\usepackage{layout}
\usepackage{mathtools}
\usepackage{afterpage}
\usepackage{color}

\begin{document}

\preprint{APS/123-QED}

\title{Electronic transport properties of titanium nitride grown by molecular beam epitaxy}

\author{Kosuke Takiguchi}
 \email{kosuke.takiguchi@ntt.com}
 \affiliation{%
NTT Basic Research Laboratories, NTT Corporation, 3-1 Morinosato-Wakamiya, Atsugi, Kanagawa 243-0198, Japan
}%
\author{Yoshiharu Krockenberger}
 \affiliation{%
NTT Basic Research Laboratories, NTT Corporation, 3-1 Morinosato-Wakamiya, Atsugi, Kanagawa 243-0198, Japan
}%
\author{Tom Ichibha}
 \affiliation{%
School of Information Science, JAIST, Asahidai 1-1, Nomi, Ishikawa 923-1292, Japan.
}%
\author{Kenta Hongo}
 \affiliation{%
School of Information Science, JAIST, Asahidai 1-1, Nomi, Ishikawa 923-1292, Japan.
}%
\author{Ryo Maezono}
 \affiliation{%
School of Information Science, JAIST, Asahidai 1-1, Nomi, Ishikawa 923-1292, Japan.
}%
\author{Yoshitaka Taniyasu}
 \affiliation{%
NTT Basic Research Laboratories, NTT Corporation, 3-1 Morinosato-Wakamiya, Atsugi, Kanagawa 243-0198, Japan
}%
\author{Hideki Yamamoto}
\affiliation{%
NTT Basic Research Laboratories, NTT Corporation, 3-1 Morinosato-Wakamiya, Atsugi, Kanagawa 243-0198, Japan
}%

\date{\today}

\begin{abstract}
    This study investigates the molecular beam epitaxial (MBE) growth of titanium nitride (TiN) thin films, achieving a high residual resistivity ratio (RRR) of 15.8. We observed a strong correlation between growth temperature and crystalline quality, as reflected in both RRR values and lattice parameter variations. Characterization of superconductivity yielded a Ginzburg-Landau coherence length of 60.4 $\pm$ 0.6\,nm, significantly higher than typical sputtered films, suggesting improved superconducting coherence. First-principles calculations, in conjunction with experimental data, provided detailed insights into the electronic structure and transport properties of the TiN films. Temperature-dependent Hall coefficient measurements further revealed the influence of anisotropic scattering mechanisms. These findings establish a promising route for the development of nitride-based superconducting materials for advanced quantum computing technologies.
\end{abstract}

\maketitle


    %

    \newpage

    \section{Introduction}
    Titanium nitride (TiN) is one of the superconducting refractory metal nitrides that has been studied for over half a century due to its rich physical properties \cite{Boltasseva2015,Guler2015}. Its high chemical stability makes TiN a promising candidate material for superconducting devices with outstanding mechanical hardness and a bulk superconducting transition temperature $T_{\rm c}$ of 5.6\,K. The observed high thermal stability of titanium nitride, in comparison to transition metal oxides, arises from the increased covalency of its chemical bonds. The observed mechanical hardness of 20\,GPa, a notable property, is a consequence of the relatively strong covalent bonding within the material \cite{Ljungcrantz1996}. TiN has been employed in quantum circuits to establish qubit systems characterized by high kinetic inductance and low-frequency noise \cite{Vissers2010,Shearrow2018,Chang2013}. Consequently, understanding the material properties of TiN is crucial for its further application.\par
    However, achieving stoichiometric TiN remains a significant challenge \cite{Herzig1987,Spengler1978,Yuan2013,Forno2019}. Nonstoichiometric TiN frequently exhibits various crystalline phases depending on the defect concentration. These disorders and defects significantly influence the electronic properties of TiN. Thus, its intrinsic physical properties remain unclear, as the preparation of stoichiometric TiN has not yet been fully realized. This issue is also reflected in the limited lifetime of transmon qubits made of TiN, which is currently around 300\,$\mu$s \cite{Deng2023}. The accurate determination of intrinsic electronic properties is impeded by the presence of defects. Consequently, minimizing defect concentrations is imperative to elucidate the material's fundamental electronic characteristics.\par
    Molecular beam epitaxy (MBE) has emerged as a highly effective technique to minimize defect concentrations. In this study, we utilized MBE to grow TiN thin films with a residual resistivity ratio (RRR) of up to 15.8. Such high RRR values emphasize effective suppression of defect scattering events. After carefully optimizing the growth temperature, we achieved epitaxial TiN films coherently grown on MgO substrates. Using the TiN thin films with RRR = 15.8, we investigated their transport properties and estimated their superconducting characteristics, complementing experimental results with first-principles calculations. Density functional theory (DFT) combined with Boltzmann transport equation analysis enabled a direct comparison between the experimentally obtained Hall coefficient data and theoretical predictions. The results provide valuable insights for the development of TiN-based superconducting devices.
    
  \section{Methods}
  \subsection{Experimental details}
  TiN films were synthesized on MgO (001) substrates using MBE. The lattice constant of the MgO substrate ($a = 4.214\,\rm{\AA}$) closely matches that of TiN ($a = 4.235\,\rm{\AA}$) \cite{Yen1967}, resulting in small lattice mismatch ($\sim$ 0.5\%), a prerequisite to sustain atomically sharp interfaces. Our custom-designed MBE system is equipped with e-beam evaporators for generating Ti flux. Controlling the Ti flux using e-beam evaporators is challenging due to titanium's high melting point (1668$^{\circ}$C). However, we achieved precise flux control through electron impact emission spectroscopy (EIES) \cite{Yamamoto2013}. The growth rate was maintained at 0.035\,nm/s, and the resulting film thickness was 42\,nm.\par
  To create a strong nitriding environment, a custom-designed atomic nitrogen source was operated at 13.56\,MHz (radio frequency, RF) with an RF power of 400\,W. The nitrogen flow rate was set to 1.0\,sccm, corresponding to a background nitrogen pressure of 4$\times$10$^{-6}$ Torr. The substrate temperature $T_{\rm{s}}$ during growth was monitored using a radiation pyrometer (Japan Sensor). After the growth, the RF power was turned off, and the film was cooled to below $T_{\rm{s}} = 200^{\circ}$C under a continuous nitrogen gas flow.\par
  We performed x-ray diffraction (XRD) and scanning transmission microscopy (STEM) for the determination of the crystal structure. We fabricated Hall-bar devices (180\,$\rm{\mu m} \times$ 300\,$\rm{\mu m}$) with 40\,nm thick silver electrodes. Longitudinal resistivity $\mathrm{\rho}_{xx}$ measurements were performed using a standard four-probe method. Magnetoresistivity and Hall resistivity $\mathrm{\rho}_{xy}$ were measured in a Quantum Design Dynacool PPMS under magnetic fields applied perpendicular to the film surface.
  \subsection{Computational Details}
  Density functional theory (DFT) calculations were performed using the generalized gradient approximation (GGA) with the Perdew-Burke-Ernzerho for solids (PBEsol) exchange-correlation functional \cite{Perdew2008}, implemented in the Quantum Espresso package \cite{Giannozzi2009}. Projector augmented wave (PAW) pseudopotentials were used, with wave function and charge density cutoffs set at 80 Ry and 960 Ry, respectively. Occupancies were determined using a Gaussian smearing approach with a broadening of 0.1 Ry. $10 \times 10 \times 10$ and $50 \times 50 \times 50$ $k$-point meshes were employed for electronic structure calculation and Boltzmann equation analysis using BoltzTraP2 \cite{Madsen2006,Georg2018}, respectively. We confirmed that the total energy convergence with respect to cutoff energy and $k$-mesh density was within $\pm 0.001$ Ry (see \cite{Sup}). For BoltzTraP2 calculations, a denser $k$-mesh is necessary to accurately simulate transport coefficients\cite{Madsen2006}. We tuned the deviation of the Hall coefficient to be with a deviation of $\pm$2.0$\times$10$^{-12}$ m$^3$/C using the $50 \times 50 \times 50$ $k$-point mesh (see \cite{Sup}). FermiSurfer \cite{Kawamura2019} is used to visualize the Fermi surfaces. The lattice parameters of TiN were set as $a = b = 4.214\,\rm{\AA}$ and $c = 4.248\,\rm{\AA}$. While bulk TiN exhibits cubic symmetry ($Fm\bar{3}m$), the compressive strain imposed by the MgO substrate induces tetragonal symmetry ($I4/mmm$) in the thin films. These lattice parameters, with $c$ estimated from our sample with the best RRR, and $a (= b)$ assumed to match the MgO substrate, enable a reasonable comparison between experimental and theoretical results.

  \section{Results and discussion}
  \subsection{Growth temperature dependence of RRR and lattice constant}
\begin{figure}[ht]
    \includegraphics[width=\columnwidth]{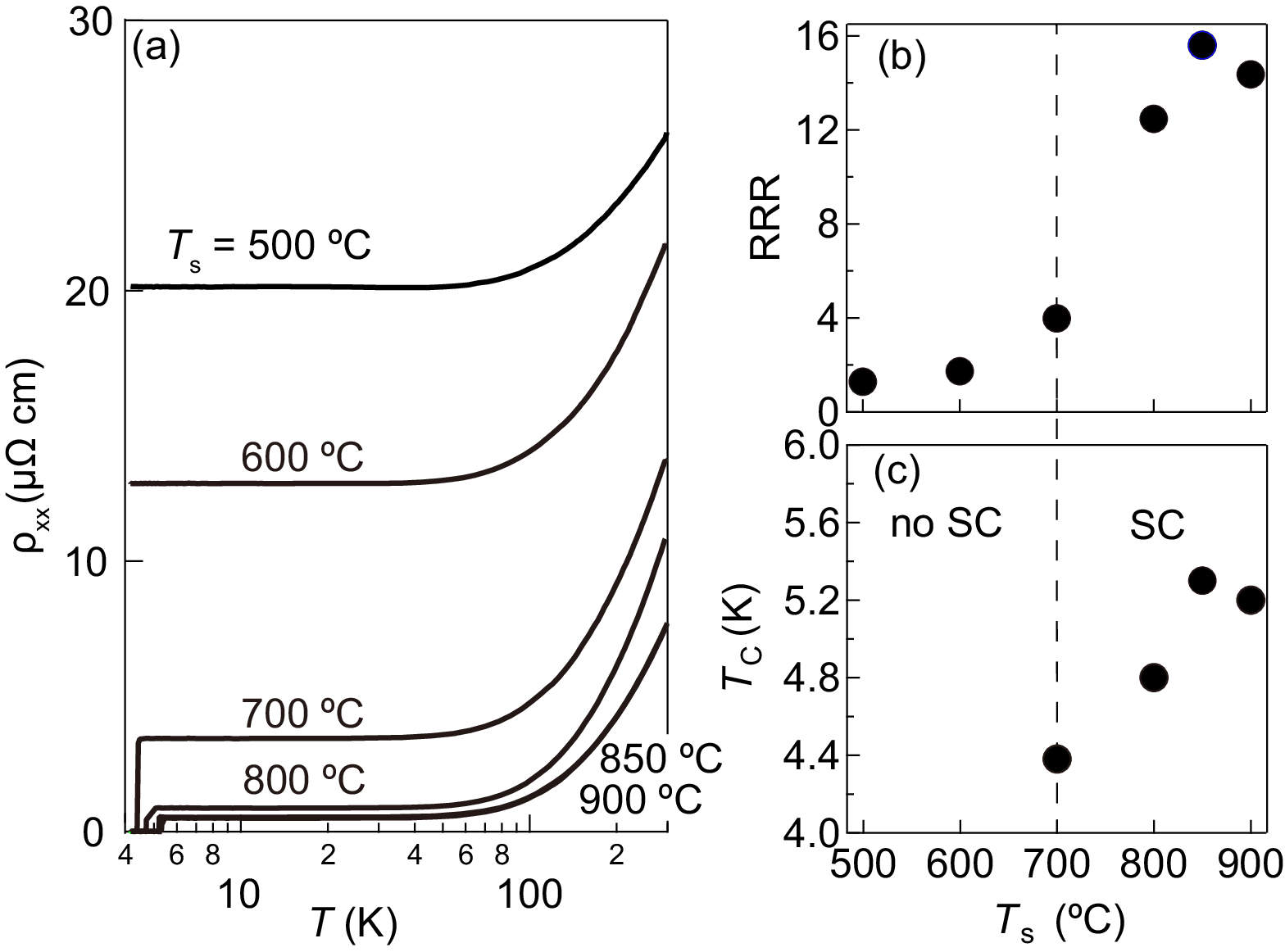}
    \caption{(a)Temperature dependence of electrical resistivity of the samples grown at various $T_{\rm{s}}$. The curves for $T_{\rm{s}} = 850\,^\circ\rm{C}$ and $900\,^\circ\rm{C}$ are almost overlapped. (b)(c) $T_{\rm{s}}$ dependence of RRR and $T_{\rm{c}}$. When $T_{\rm{s}} < 700\,^\circ\rm{C}$, no superconductivity behaviour is seen down to 4.2\,K.}
    \label{fig:rhoT}
  \end{figure}

  The choice of growth parameters is critical for achieving high-quality thin films. A higher $T_{\rm{s}}$ is more suitable for metal nitride growth compared to oxide growth due to the relatively slower kinetic activity between metal and nitrogen ions. Under thermodynamic equilibrium conditions, stoichiometric TiN is inaccessible for temperatures below 1000$^\circ$C \cite{Vahlas1991}. This tendency also holds for thin film synthesis using atomic nitrogen as a reactant. These characteristics are also observed in our samples. We synthesized TiN thin films at various substrate temperatures: $T_{\rm{s}} = 500\,^\circ \rm{C}, 600\,^\circ \rm{C}, 700\,^\circ \rm{C}, 800\,^\circ \rm{C}, 850\,^\circ \rm{C}$, and $900\,^\circ \rm{C}$. Figure \ref{fig:rhoT}(a) illustrates the temperature dependence of resistivity for samples grown at different $T_{\rm{s}}$. Superconductivity is induced if $T_{\rm{s}} \geq 700\,^\circ\rm{C}$. Furthermore, RRR and $T_{\rm{C}}$ increase monotonically for $T_{\rm{s}} \leq 850\,^\circ\rm{C}$.\par
  The RRR value of 15.8, which is the highest reported to date \cite{Krockenberger2012_,Inumaru2000,Vissers2013,Bi2021,Guo2019,Richardson2020,Femi-Oyetoro2024,Zhang2024}, was obtained at $T_{\rm{s}} = 850\,^\circ\rm{C}$. However, at $T_{\rm{s}} = 900\,^\circ\rm{C}$, both the RRR and $T_{\rm{c}}$ decreased compared to their maximum values at $T_{\rm{s}} = 850\,^\circ\rm{C}$. For $T_{\rm{s}} > 900\,^\circ\rm{C}$, TiN dissociates into off-stoichiometric phases, as previously reported \cite{Vahlas1991}.\par
  
\begin{figure}[ht]
    \includegraphics[width=\columnwidth]{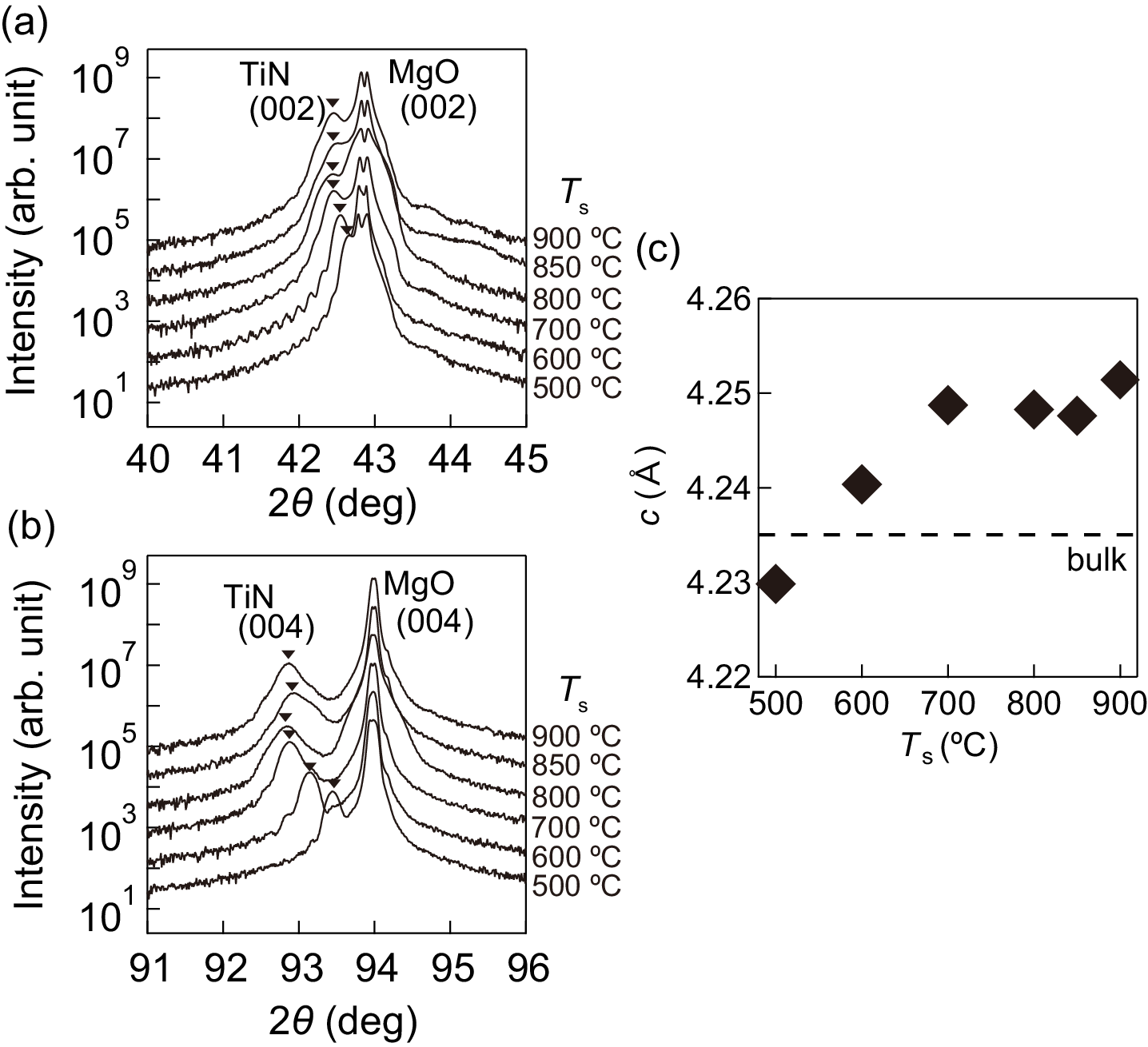}
    \caption{X-ray diffraction spectra around (a) (002) and (b) (004) peaks of TiN grown at various $T_{\rm{s}} ( = 500\,^\circ \rm{C}, 600\,^\circ \rm{C}, 700\,^\circ \rm{C}, 800\,^\circ \rm{C}, 850\,^\circ \rm{C},$ and $ 900\,^\circ \rm{C})$.(c) $T_{\rm{s}}$ dependence of $c$ axis length of the TiN thin films. The dashed line indicates the lattice constant of bulk TiN.}
    \label{fig:xrd}
  \end{figure}

  The structural investigation provides further insight into the growth temperature dependence of TiN thin films. Figures \ref{fig:xrd}(a) and (b) present the XRD patterns of TiN thin films grown at various $T_{\rm{s}}$. The (002) and (004) diffraction peaks of TiN are clearly visible in all samples, with no peaks corresponding to other crystalline phases, confirming the successful epitaxial growth of the films on MgO (001) substrates. (See also \cite{Sup}.) The systematic peak shifts with increasing $T_{\rm{s}}$ indicate changes in the lattice constant due to variations in nitrogen deficiency.\par
  Figure \ref{fig:xrd}(c) shows the variation of $c$ as a function of $T_{\rm{s}}$. The value of $c$ is determined using the Nelson-Riley analysis \cite{Nelson1945} based on the positions of the (002) and (004) peaks in each sample. When $T_{\rm{s}} < 700\,^\circ\rm{C}$, where no superconducting behavior is observed, $c$ increases with $T_{\rm{s}}$. This expansion of $c$ at lower growth temperatures is attributed to nitrogen deficiency \cite{Krockenberger2012_}. In contrast, when $T_{\rm{s}} > 700\,^\circ\rm{C}$, $c$ remains nearly constant ($4.248\,\rm{\AA}$$<c<$$4.251\,\rm{\AA}$), indicating that nitrogen deficiency is sufficiently suppressed at these higher temperatures.\par
  The reference bulk lattice constant of TiN is reported as 4.235\,\AA \cite{Yen1967}. We observed that the $c$ lattice constants in our TiN films, with the exception of the sample grown at $T_{\rm{s}}$ = 500\,$^\circ$C, were consistently larger than this bulk value. This discrepancy can be explained by the presence of compressive strain exerted by the MgO substrate, which possesses a lattice constant of $a$ = 4.214\,\AA. The strain effect, visualized in Fig. 2(c), provides additional evidence for the successful epitaxial growth of TiN on the MgO substrate.\par
\begin{figure}[ht]
  \includegraphics[width=0.8\columnwidth]{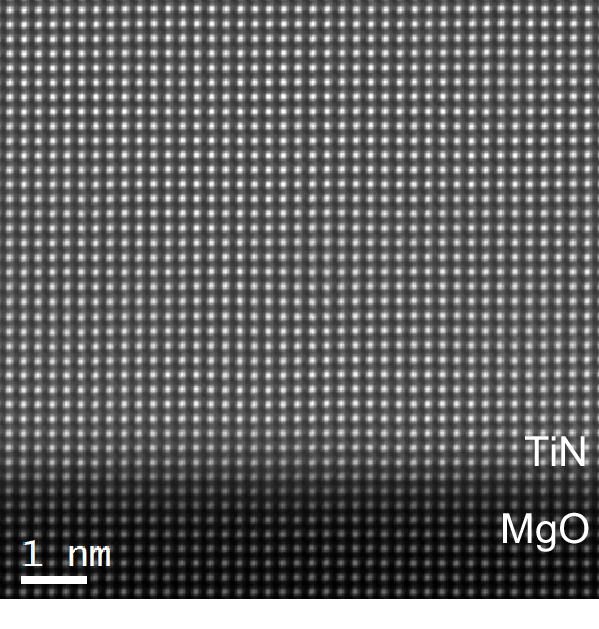}
  \caption{HAADF-STEM image of the TiN film grown at $T_{\rm{s}}=850\,^\circ\rm{C}$. Electron incident direction is in [100] of the MgO substrate. The scale bar corresponds to 1\,nm.}
  \label{fig:stem}
  \end{figure}

  High-angular annular dark field scanning transmission electron microscopy (HAADF-STEM) images clearly show the epitaxial growth of our TiN thin film. A HAADF-STEM image of the TiN film with the highest RRR is displayed in Figure \ref{fig:stem}. The Ti atomic columns are imaged with a sharpness comparable to the Mg columns of the substrate, demonstrating substantial in-plane crystalline coherence throughout the film. The image reveals a notable lack of dislocations and other defects, which is consistent with the measured high RRR value.

  \subsection{Upper critical field analysis for the epitaxial TiN thin film}
\begin{figure}[ht]

    \includegraphics[width=\columnwidth]{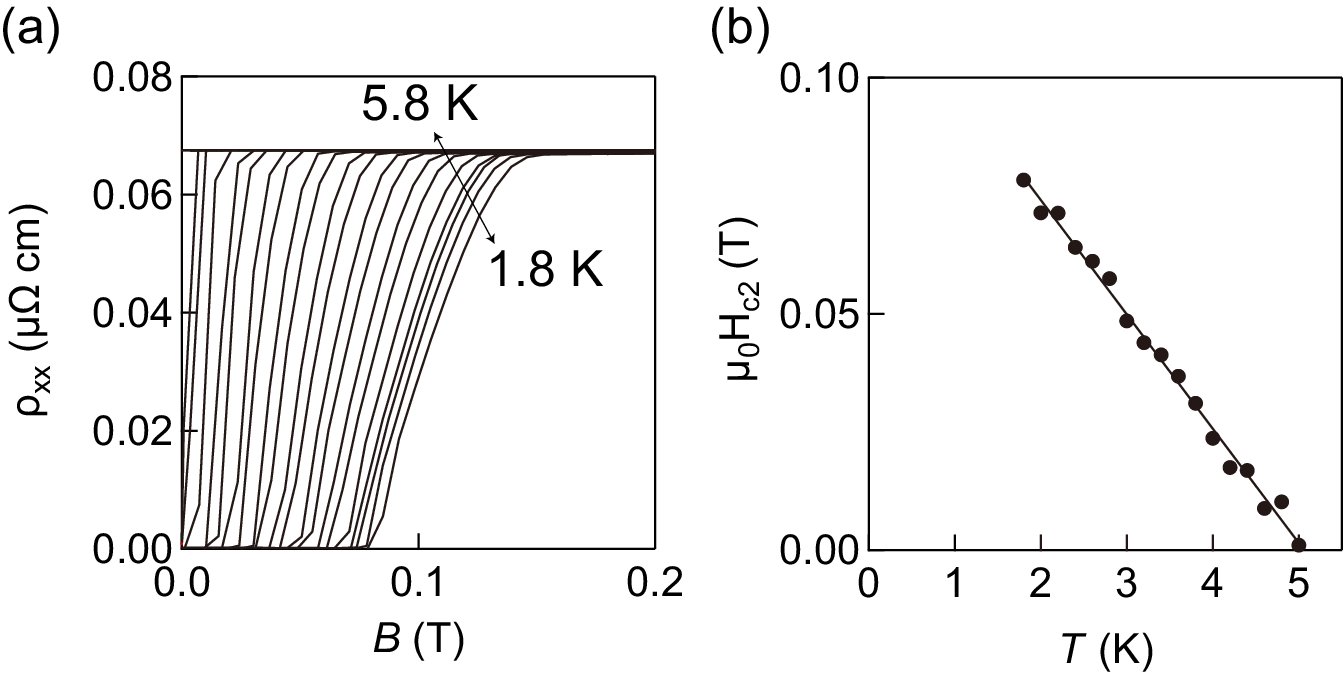}
      \caption{(a)Magnetic field dependence of the resistivity. Temperature range is from 1.8\,K to 5.8\,K with increment of 0.2\,K. (b) Temperature dependence of the upper critical field. According to GL theory, we performed the linear fitting to estimate $H_{\rm{c2}}(0)$.}
      \label{fig:Hc2}

    \end{figure}

  The intrinsic physical properties of TiN, including key superconducting parameters such as the upper critical field ($H_{\rm{c2}}$) and coherence length, can be accurately determined using the sample with an RRR of 15.8. These parameters are critical for the design and development of quantum devices. Figure \ref{fig:Hc2}(a) presents the magnetoresistivity data measured from 1.8\,K to 5.8\,K. A clear phase transition from the superconducting to the normal state is observed, enabling the extraction of $H_{\rm{c2}}$ as a function of temperature. Near $T \simeq T_{\rm{c}}$, $H_{\rm{c2}}$ exhibits a linear dependence on temperature. This is consistent with predictions from Ginzburg-Landau (GL) theory in $s$-wave superconductors. Although the strain effect breaks the cubic crystalline symmetry of TiN, the superconducting order parameter in our TiN film is isotropic. The GL coherence length $\xi_{\rm{GL}}$ may be estimated by
  \begin{align}
      \xi_{\rm{GL}} &= \sqrt{\frac{\mathit{\Phi}_0}{2\pi H_{\rm{c2}}(0)}},\\
      H_{\rm{c2}}(0) &= 0.69T_{\rm{c}} \left|\frac{dH_{\rm{c2}}(T)}{dT}\right|_{T=T_{\rm{c}}},
  \end{align}
  where $\mathit{\Phi}_0$ is a magnetic flux quanta. The estimated $\xi_{\rm{GL}}$ is 60.4 $\pm$ 0.6\,nm (see also Table \ref{tab:exp}.), and the reported ones in TiN films are 2.0 - 39\,nm \cite{Draher2023,Zhang2024,Krockenberger2012_}. The enhancement of $\xi_{\rm{GL}}$ achieved through our MBE method is consistent with the fact that the suppression of defects and impurities is greater than that observed in other growth techniques, such as sputtering. 
 
    \begin{table}[ht]

      \caption{Physical parameters experimentally measured in the TiN film. $T_{\rm{c}}$ is the temperature at $\rho_{xx}(T)=0$. $\rho_{\rm{const}}$ is defined as $\rho_{xx}(\rm{6.0\,K})$. \label{tab:exp}}
      \begin{tabular}{l|c}
        \hline \hline
        RRR & 15.8 \\
        $T_{\rm{c}}$ (K) & 5.30 $\pm$ 0.01 \\
        $\rho_{\rm{const}}$ ($\mu \Omega$\,cm) & 0.49 $\pm$ 0.01  \\
        $H_{\rm{c2}}(0)$ (mT) & 90 $\pm$ 2.0 \\
        $\xi_{\rm{GL}}$ (nm)& 60.4 $\pm$ 0.6\\
        \hline \hline
      \end{tabular}

    \end{table}

    \begin{table}[ht]

      \caption{Physical parameters estimated by using the DFT calculation results. \label{tab:dft}}
      \begin{tabular}{l|c}
        \hline \hline
        $v_\mathrm{F}$ (m/s) & 1.87 $\times$ 10$^6$ \\
        $k_F$ (m$^{-1}$) & 1.62 $\times$ 10$^{10}$ \\
        $n$ (m$^{-3}$) & 4.60 $\times$ 10$^{28}$ \\
        $\ell$ (nm) & 29.4 $\pm$ 0.4 \\
        $\xi_0$ (nm)& 48.5 $\pm$ 0.1\\
        $\xi$ (nm)& 18.3 $\pm$ 0.2\\
        \hline \hline
      \end{tabular}

    \end{table}
    \subsection{Electronic structure and Hall effect}
\begin{figure}[ht]
      \includegraphics[width=\columnwidth]{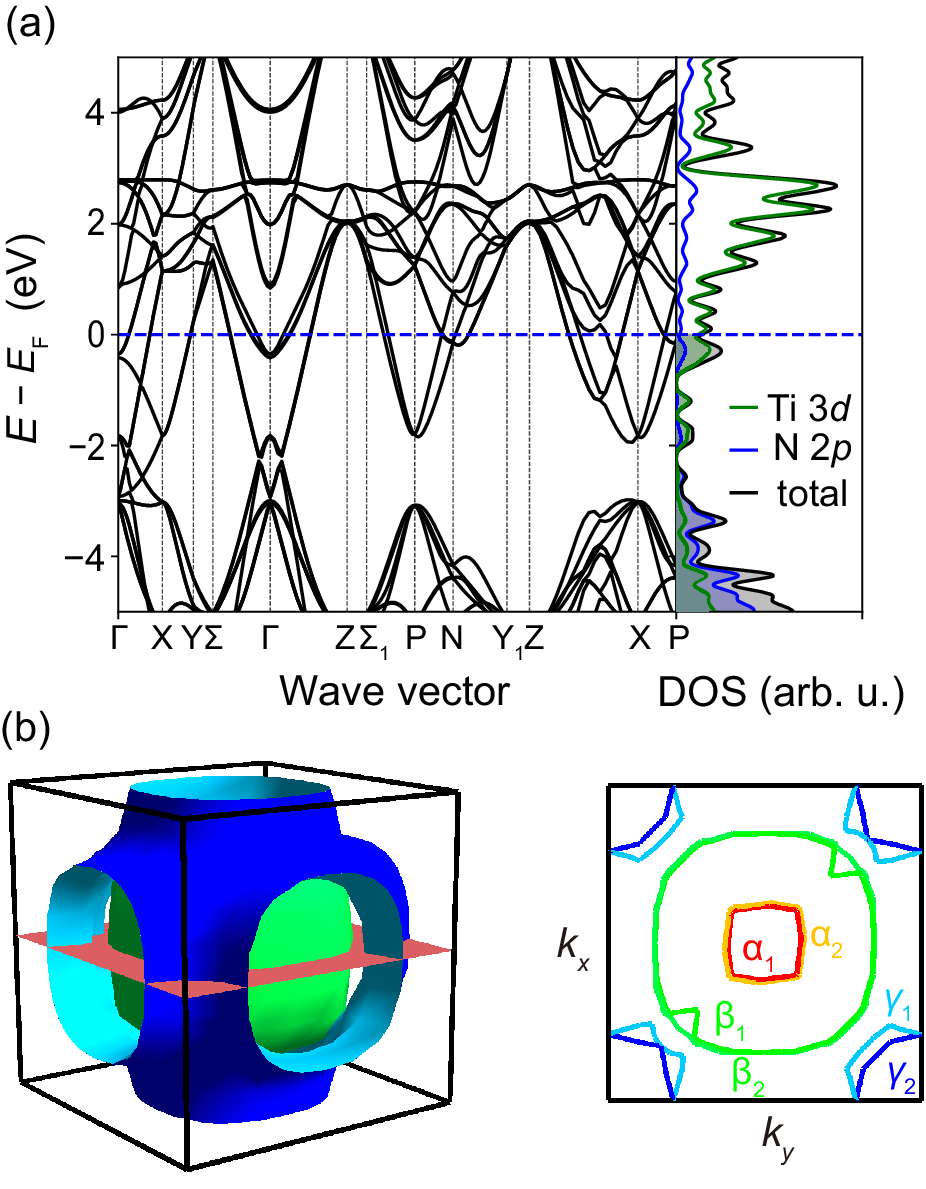}
      \caption{(a)DFT calculation results of energy dispersion and density of states (DOS) in TiN. 86\% of the total DOS consists of Ti 3$d$ orbit. (b)(left) Three dimensional Fermi surfaces of TiN obtained by DFT calculation. (right) Cross-sectional view of the Fermi surfaces of the (001) plane. The section is shown in the left panel as the pink plane. The center points in both figures correspond to $\Gamma$ point. We assigned $\alpha_1$, $\alpha_2$, $\beta_1$, $\beta_2$, $\gamma_1$ and $\gamma_2$ to each Fermi surface. $\alpha_1$, $\alpha_2$, $\beta_1$, and $\beta_2$ Fermi surfaces are closed, and mainly contribute to the electronic conduction. }
      \label{fig:dft_el}
    \end{figure}

    To further understand the intrinsic properties of TiN, we determined first the Hall coefficient and compared it's temperature dependence to results derived by DFT calculations. Figures \ref{fig:dft_el}(a) and (b) show the DFT-calculated electronic structure and Fermi surfaces of TiN, respectively. 86\% of the total density of states (DOS) at the Fermi level $E_{\rm{F}}$ consists of Ti 3$d$ orbitals. There are six Fermi surfaces, four of which are closed and primarily contribute to electronic conduction as the main carriers. This electronic conduction is experimentally observed in our Hall measurements (see \cite{Sup} for $\rho_{xy}(B)$ curves.) and previous studies \cite{Zhang2024}.\par
    However, interpreting the Hall effect in the presence of such a variety of Fermi surfaces is not straightforward. A semi-classical approach to transport coefficients, based on the Boltzmann equation, provides a quantitative framework for discussion. Using BoltzTraP2 \cite{Madsen2006,Georg2018}, we calculated the Hall coefficient $R_{\rm{H}}$ by incorporating group velocities derived from the DFT results. Importantly, the BoltzTraP2 calculations rely on the constant relaxation time approximation, meaning the relaxation time $\tau$ is assumed to be isotropic. Comparing our experimental data with the calculations enables discussion of the anisotropy in $\tau$.\par
\begin{figure}[ht]
      \includegraphics[width=\columnwidth]{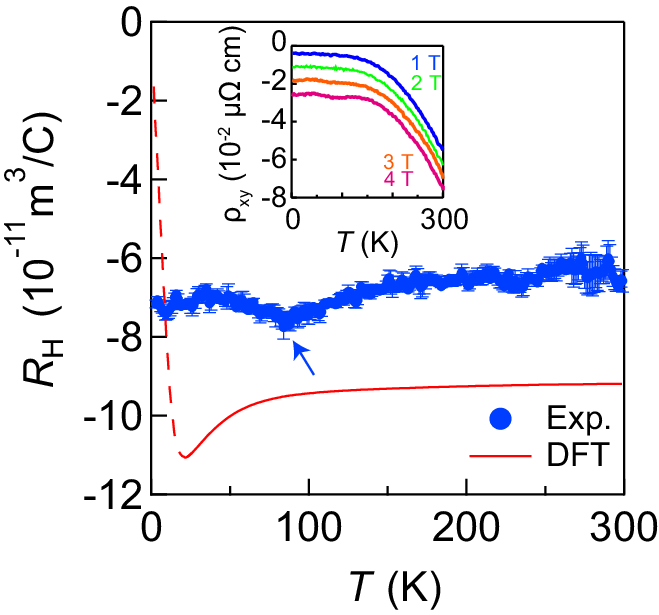}
      \caption{Temperature dependence of Hall coefficient $R_{\rm{H}}$ estimated by the experimental measurements (blue) and the DFT calculation (red). The abrupt jump in the calculation result (red dashed line) is due to the numerical instability near 0\,K of $\partial f/\partial E$, where $f$ is Fermi-Dirac distribution function. The inset shows $\rho_{xy}(T)$ curves in the magnetic field $B= $ 1\,T, 2\,T, 3\,T, and 4\,T. The kink at $T=84$\,K in the experimental data pointed by the arrow. }
      \label{fig:hall_T}
    \end{figure}

    Figure \ref{fig:hall_T} shows the temperature dependence of the Hall coefficient $R_{\rm{H}}$ determined experimentally and by calculations. Experimental $R_{\rm{H}}$ are estimated by calculating $d\rho_{xy}(B)/dB$ in $B=$ 1\,T, 2\,T, 3\,T, and 4\,T as shown in the inset. If 84\,K $< T <$ 300\,K, the experimental value of $|R_{\rm{H}}|$ increases as temperature decreases. At $T = 84\,\rm{K}$, the $|R_{\rm{H}}|$ curve exhibits a kink and subsequently decreases with decreasing temperature. The temperature dependence of $|R_{\rm{H}}|$ can be attributed to anisotropic electron-phonon scattering involving Umklapp processes \cite{Ziman1961,Hurd1972}. In the case of TiN, such scattering appears to be suppressed in the range of 84\,K $< T <$ 300\,K, resulting in an increase in $|R_{\rm{H}}|$ by decreasing temperature. The calculated $|R_{\rm{H}}|$ values overestimate the experimental results, likely due to the absence of anisotropic phonon scattering in the calculations and depolarization. In the calculations of $R_{\rm{H}}$, the kink structure at 20\,K and the abrupt change in $T<20\,\rm{K}$ (dashed red line) is attributed to the numerical instabilities in $\partial f/\partial E$ near $T=0\,$K, where $f (=[1+\exp((E-E_F)/(k_\mathrm{B}T))]^{-1})$ is Fermi-Dirac distribution function. \par 
    By determining the upper limit of RRR through calculations, a deeper understanding can be gained regarding the extent to which impurities and defects in thin films affect electrical conduction. $\tau$ from electron-electron scattering can be expressed as
      \begin{align}
        \frac{\hbar}{\tau}\simeq\frac{(k_\mathrm{B}T)^2}{E_{\rm{F}}},
      \end{align}
    where $k_\mathrm{B}$ is Boltzmann constant, and $E_{\rm{F}}$ is Fermi energy \cite{AshcroftMermin}. Based on this analysis, the upper limit of RRR of TiN is theoretically estimated to be 4458. (See also \cite{Sup} for the simulated $\rho_{xx}(T)$ curve.) The significant deviation from the experimentally measured RRR value of 15.8 highlights the critical role of disorders in determining the physical properties. As noted in earlier studies on bulk TiN \cite{Herzig1987,Spengler1978}, such disorders are inevitable, even in stoichiometric samples. This suggests that imperfections in the crystal structure are key factors in understanding the material's behavior. \par
    
    To gain quantitative insights into the effects of impurities and disorders, estimating the mean free path $\ell$ is also valuable. Applying the free electron model to our TiN film, the Fermi velocity $v_\mathrm{F}$ and carrier concentration $n$ are estimated by the DFT results and are summarized in Table \ref{tab:dft}. It is important to note that estimating $n$ from Hall effect data is not straightforward for TiN. TiN has six Fermi surfaces, making it impossible to directly convert $R_{\rm{H}}$ data into $n$. The estimated values of $v_\mathrm{F} (= 1.87 \times 10^6$\,m/s) and $n (= 4.60 \times 10^{28}\,\rm{m^{-3}}$) are comparable to previously reported data ($v_\mathrm{F} = 7 \times 10^5$\,m/s, $n = 5.3 \times 10^{28}\,\rm{m^{-3}}$) \cite{Chawla2013}. The mean free path $\ell$ is calculated using the following equation 
      \begin{align}
        \ell=\frac{mv_\mathrm{F}}{e^2n\rho_{\rm const}},
      \end{align}
    where $m$ is electron mass, $e$ is elemental charge, and $\rho_{\rm const}$ is resistivity derived from our experimental data. Based on this analysis, $\ell$ is estimated to be 29.4 $\pm$ 0.4\,nm. Additionally, according to GL theory, BCS superconducting coherence length $\xi_0$ can be estimated as
      \begin{align}
        \xi_0=\frac{0.18\hbar v_\mathrm{F}}{k_\mathrm{B}T_{\rm{c}}}.
      \end{align}
    $\xi_0$ is estimated to be 48.5 $\pm$ 0.1\,nm. Pippard coherence length is $\xi = (1/\xi_0 + 1/\ell)^{-1}=18.3\pm 0.2$\,nm. Since $\xi_0$ and $\xi$ are of the same order of magnitude, the system approaches the clean limit condition ($\xi\sim\xi_0$). These parameters are crucial for the design and optimization of TiN-based quantum devices.

  \section{Conclusion}
  We investigated the MBE growth of epitaxial TiN thin films at various growth temperatures and successfully obtained a film with an RRR of 15.8. The growth temperature significantly influences the crystalline quality, as evidenced by the RRR and the variation in lattice constants. Lower growth temperatures induce nitrogen deficiency in the TiN films, leading to off-stoichiometric compositions.\par
  Using the TiN thin film with an RRR of 15.8, we performed magnetotransport measurements. The Ginzburg-Landau (GL) coherence length, derived from upper critical field measurements, was estimated to be 60.4 $\pm$ 0.6\,nm. This indicates longer superconducting coherence in the MBE-grown sample compared to sputtered films \cite{Draher2023,Zhang2024}. Furthermore, a comparison of theoretical calculations and experimental results for the temperature dependence of the Hall coefficient provided insights into the electronic structure of TiN. The discrepancies between theory and experiment were attributed to anisotropic electron phonon scattering. The estimated BCS and Pippard coherence lengths being of the same order of magnitude suggest that the system largely satisfies the clean limit condition. However, extrinsic factors such as disorder effects remain key to understanding the practical properties of TiN.\par
  As discussed, epitaxial growth techniques are essential for uncovering the intrinsic properties of materials. These findings offer a pathway to utilizing nitride superconducting materials.

  \section*{Acknowledgment}
  The authors thank T. Ikeda for his support in high-angle annular dark field scanning transmission electron microscope (HAADF-STEM) measurements. The computation in this work has been performed using the facilities of the the Research Center for Advanced Computing Infrastructure (RCACI) in JAIST. T. I. appreciates the support from JSPS KAKENHI Grant number 24K17618 and JSPS Overseas Research Fellowships.

    \section*{References}
    \renewcommand{\bibsection}{}

    \bibliography{PrReN}


\end{document}